\documentclass[10pt, conference]{IEEEtran}

\usepackage{amsmath,amssymb}
\usepackage[linesnumbered,ruled]{algorithm2e}
\usepackage{graphicx,xcolor,array}
\usepackage{pgfplots}
\pgfplotsset{compat=1.14}
\usepackage{tikz,cite}
\usetikzlibrary{arrows}
\usetikzlibrary{automata,positioning}
\usepackage{makecell}
\usepackage{upgreek}
\usepackage{pifont}
\usepackage{mathscinet}
\usepackage[letterpaper, left=0.625in,right=0.625in,top=0.75in,bottom=1in]{geometry}
\usepackage{geometry}

\usepackage[colorlinks=true,linkcolor=black,anchorcolor=black,citecolor=black,filecolor=black,menucolor=black,runcolor=black,bookmarks=false,hidelinks,urlcolor=black]{hyperref}
\usepackage{enumitem}

\newcommand{\seb}[1]{\textcolor{black}{#1}}

\newif\ifcomment
\commenttrue
\newcommand{\np}[1]{\textcolor{black}{#1}}

\newtheorem{definition}{Definition}

\renewcommand{\baselinestretch}{0.974}

\makeatletter
\def\endthebibliography{%
  \def\@noitemerr{\@latex@warning{Empty `thebibliography' environment}}%
  \endlist
}
\makeatother

\begin{document}

\title{Fairness Notions in DAG-based DLTs}

\author{
\IEEEauthorblockN{
Mayank Raikwar,\IEEEauthorrefmark{1} Nikita Polyanskii,\IEEEauthorrefmark{2} Sebastian M{\"u}ller\IEEEauthorrefmark{3}
\IEEEauthorblockA{\IEEEauthorrefmark{1}University of Oslo, Norway\\
Email: mayankr@ifi.uio.no}
\IEEEauthorblockA{\IEEEauthorrefmark{2}IOTA Foundation, Berlin, Germany\\
Email: nikita.polianskii@iota.org}
\IEEEauthorblockA{\IEEEauthorrefmark{3}Aix Marseille Universit{\'e}, CNRS, Centrale Marseille, France\\
Email: sebastian.muller@univ-amu.fr}
    }}


\maketitle

\begin{abstract}
This paper investigates the issue of fairness in Distributed Ledger Technology (DLT), specifically focusing on the shortcomings observed in current blockchain systems due to Miner Extractable Value (MEV) phenomena and systemic centralization. We explore the potential of Directed Acyclic Graphs (DAGs) as a solution to address or mitigate these fairness concerns.
Our objective is to gain a comprehensive understanding of fairness in DAG-based DLTs by examining its different aspects and measurement metrics. We aim to establish a shared knowledge base that facilitates accurate fairness assessment and allows for an evaluation of whether DAG-based DLTs offer a more equitable design.
We describe the various dimensions of fairness and conduct a comparative analysis to examine how they relate to different components of DLTs. This analysis serves as a catalyst for further research, encouraging the development of cryptographic systems that promote fairness.
\end{abstract}

\section{Introduction}
DLTs have become quite popular in the last decade due to their broad range of applications. The most well-known DLT architecture is the Blockchain which was introduced in Bitcoin cryptocurrency~\cite{nakamoto2008bitcoin}. After the advent of Bitcoin, there have been a plethora of cryptocurrencies~\cite{tschorsch2016bitcoin} with similar underlying blockchain structures but differing in consensus, security guarantees, performance, and additional functionalities. Nevertheless, the main technical hurdle to the adoption of blockchain in real-world applications is its scalability. Therefore, many methods such as Layer-2 protocols~\cite{gudgeon2020sok}, and sharding techniques~\cite{wang2019sok} have been proposed to solve the scalability issue of the blockchain. Another alternative solution to the scalability problem is to have a different structure of DLT than the linear chain structure of blockchain. One such structure is Directed-Acyclic Graph (DAG)-based DLT~\cite{sompolinsky2013accelerating,popov2018tangle}.
The assurance of DAG-based DLT is high scalability and fast confirmation of transactions without significantly compromising security.

The idea of DAG-based DLT is to form a directed acyclic graph structure from the blocks/transactions instead of forming a chain by referencing a single block~\cite{wang2023sok}. In this way, many transactions are verified in parallel and hence provide a more scalable DLT architecture. The goal of these DAG-based DLTs is to \np{achieve higher throughput while keeping or reducing the communication or/and computation complexity, latency, and storage overhead.} 
Most of the existing works in DAG-based DLTs focus on performance improvement, however, fairness notions in DAG-based DLTs have not yet been explored.

Fairness plays a prominent role in DLTs encompassing technical, economic, and social considerations. In a DLT, an unfair scenario can lead to discontent among participants, thereby impeding the widespread adoption of the DLT. Consequently, it is imperative for DAG-based systems to prioritize the integration of fairness within their frameworks. The prevailing notion of fairness, often acknowledged, entails the eventual acceptance of blocks produced by slow yet honest nodes in the DAG-based DLT. A node can be termed as \textit{slow} if 1) the node has low computing power in Proof-of-Work (PoW)-based DLT; 2) the node has a low stake in Proof-of-Stake (PoS)-based DLT; 3) the node has large communication delay with the rest of the network, for example, due to the network topology or being geographically distant. The above-defined notion is a very general definition of fairness but in DAG-based DLTs, fairness can be expressed in different notions.

While fairness has been examined and addressed in blockchain protocols, the presence of a single miner or validator making decisions regarding the ledger's progress remains a prominent cause of fairness challenges. However, in DAG-based protocols, the ledger's state progression is controlled by multiple nodes. In essence, this represents the potential for enhanced fairness in DAG-based protocols. Furthermore, some DAG-based DLTs allow weak references to rescue the old blocks and improve the fairness of the DLT~\cite{spiegelman2022bullshark, popov2020coordicide}.

\subsection{Related Work}
Since the inception of Bitcoin, there has been a \seb{substantial} amount of work and studies conducted toward improving the scalability and security of the protocol~\cite{conti2018survey}. Nevertheless, in recent years, a few works concentrate on fairness in Bitcoin like linear blockchain models. These works examine fairness in two main aspects: 1) Fairness in reward mechanisms of blockchain systems~\cite{amoussou2019fairness,huang2021rich}; 2) Fairness in the ordering of transactions in blockchain systems~\cite{kursawe2020wendy,kelkar2020order}.

Amoussou-Guenou et al.~\cite{amoussou2019fairness} propose the formal notions of fairness in committee-based PoW blockchain protocols. The authors define two fairness notions: 1) Selection fairness which says how fairly the subset of nodes are selected to participate to append a block in a consensus round; 2) Reward fairness which describes how the reward is distributed among the committee nodes in a fair manner. Huang et al.~\cite{huang2021rich} propose two types of fairness, i.e., exceptional fairness and robust fairness in PoS blockchain protocols. These fairness notions characterize the relation between a miner's stake and his/her reward in a consensus round.

Fairness in transaction ordering refers to the ordering of transactions in a block of a blockchain. A fair ordering expresses a fair block proposal where the block proposer follows the random selection of transactions and does not include/exclude/prioritize some transactions. Different fairness notions in transaction ordering have been proposed in blockchain protocols~\cite{kursawe2020wendy,kelkar2020order}. Even so, manipulation in transaction ordering leads to MEV extraction and MEV attacks~\cite{yang2022sok}. There are different types of MEV attacks. Out of these attacks Sandwich and Long-trail MEV attacks occur more often. In 2022, MEV searchers generated \$128 million revenue alone from Sandwich attacks~\cite{sandwich}.

Though a few efforts have been made to study and subsequently solve the fairness issues in blockchain protocols, not many works endeavour to look into fairness problems in DAG-based DLTs. The primary goal of these DLTs has been to solve the inherent low throughput problem of blockchain.



The field of DAG-based protocols has experienced continuous growth in research efforts~\cite{wang2023sok}. While many of these works aim to enhance the performance of DAG-based Distributed Ledger Technologies (DLTs), they also explore different network models (synchronous, partially synchronous, asynchronous) and adversarial models (static, adaptive). However, only a limited number of works specifically address fairness in DAG-based DLTs.
Birmpas et al.~\cite{birmpas2020fairness} present a parametric model that evaluates the fairness and efficiency of DAG-based DLTs by adjusting various ledger parameters. In their model, fairness is defined as the proportional reward allocation to participating validator nodes based on their hash power. Their findings demonstrate that fairness is significantly influenced by the connectivity levels of validator nodes within the underlying peer-to-peer network. It is important to note that their model focuses on capturing fairness exclusively in DAG-based DLTs that employ Proof-of-Work (PoW) puzzles as their consensus mechanism.

DAG-based DLTs such as DAG-Rider~\cite{keidar2021all}, Tusk~\cite{danezis2022narwhal}, Bullshark~\cite{spiegelman2022bullshark} give attention to fairness in their protocol. We will give more details about this in the following sections.

In a PoW-based DLT protocol with high difficulty, slow nodes rarely get a chance to append a block or issue the transactions. Vigneri et al.~\cite{vigneri2019achieving} propose an adaptive rate control mechanism in a PoW-based DLT providing dynamic difficulty for the fair treatment of nodes irrespective of their computational power. Another work~\cite{vigneri2020fairness} in a similar direction shows fairness between low and high power nodes through analytical and simulation results. M{\"u}ller et al.~\cite{muller2021fairness} construct a mathematical model for fairness in a weighted voting consensus protocol in DAG-based DLT. Using their model, a node having low weight also participates in consensus finding. These types of fairness notions are out of the scope of this paper.

\subsection{Our Contribution}
In this paper, we are primarily interested to explore the different notions of fairness in the context of DAG-based DLTs. To the best of our knowledge, no work has been done to systematize fairness notions in DAG-based DLTs. We present informal definitions of these fairness notions and describe the importance of these notions. We also discuss the relation of fairness with other DLT components.

\subsection{Structure of the Paper}
The rest of the paper is described as follows: Section~\ref{Preliminaries} presents the description of \np{a} DAG-based DLT. Section~\ref{Fairness} demonstrates the different fairness notions in DAG-based DLT protocols. Further, Section~\ref{Discussion} provides a brief description of the relationship between defined notions and other important components/functionalities in DAG-based DLTs. Finally, in Section~\ref{Conclusion}, we conclude the paper with a few interesting research directions.

\section{DAG-based DLT} \label{Preliminaries}
DAG-based DLT systems structure the transactions/blocks in the form of a DAG topology. A directed graph $G$ is a tuple of vertex set $V$ and directed edge set $E$, such that $G := (V, E)$. In this graph, a vertex represents a unit (transaction/block) and a directed edge represents a relationship between two units. That means, an edge $e \in E$ in the graph $G$ represents a partial order relationship between a pair of units $(u,v)$, where $u,v \in V$. An edge $u \leftarrow v$ can have the meaning that $v$ verifies/confirms/witnesses $u$. An edge can also represent a hash reference from one unit to another. A directed graph $G$ is a directed acyclic graph if $G$ has no directed cycles, whereas a \textit{directed cycle} is a sequence of vertices where the first and last vertices in the sequence are the same, and all edges are oriented in the same direction. 

The introduction of multiple references per block results in the formation of a Directed Acyclic Graph (DAG). In a traditional linear blockchain, the longest chain rule serves a dual purpose: agreeing on the included blocks and establishing their order. However, when transitioning from linear chains to DAGs, these two functions need to be replaced. To address this, two primary classes of protocols have emerged: those that perform these tasks directly on the DAG itself, and Byzantine Fault Tolerant (BFT) protocols that utilize supplementary broadcast primitives.

The concept of using DAG in a ledger protocol was first introduced by Sompolinsky et al. in GHOST protocol~\cite{sompolinsky2013accelerating} which addresses the concurrency problem of transactions in Bitcoin by building a tree instead of a linear chain. Later, an improved version of the protocol~\cite{lewenberg2015inclusive} was utilized in Ethereum~\cite{wood2014ethereum}. In GHOST protocol, a newly created (mined) block does not refer to the leaf block of the longest chain. Instead, starting from the genesis block and at each fork, the branch with the heaviest sub-tree underneath is chosen until a leaf block is found. Once the leaf block is found, the newly mined block connects to it by including the hash pointer of the leaf block. Therefore, in the GHOST protocol, blocks created by the forks also contribute to the final consensus while maintaining robustness and high throughput.

Later, IOTA~\cite{popov2018tangle}, Byteball~\cite{churyumov2016byteball}, and many other DAG-based DLT protocols adopted the paradigm of using individual transactions as vertices in a DAG. With this modification, the transactions are disseminated and confirmed faster. In the following years, many works and studies on DAG-based DLTs are conducted. These works update and modify the DAG structure, and improve the performance by requiring less computation, communication, and storage. 

The structure of the DAG becomes very relevant while achieving the consensus and proving the security guarantees. Many Proof-of-Work (PoW) employed unstructured DAG-based protocols such as SPECTRE~\cite{sompolinsky2016spectre},  PHANTOM~\cite{sompolinsky2018phantom}, and DAG KNIGHT~\cite{sompolinsky2022dag} have been proposed. In protocols like SPECTRE, PHANTOM and DAG KNIGHT, a newly created block has to cross-refer (point) to all the visible tips (leaf) in the ledger. The ordering of all these blocks in the DAG becomes the heart of the consensus protocol. Every node locally runs the ordering procedure and returns a linear ordering over the blocks in its local DAG.

Structured DAG-based protocols such as Aleph~\cite{gkagol2019aleph}, DAG-Rider~ \cite{keidar2021all}, Tusk~\cite{danezis2022narwhal}, Bullshark~\cite{spiegelman2022bullshark}, Cordial Miners~\cite{keidar2022cordial} work in rounds in a permissioned setting, i.e., there are fixed $n$ nodes that can add a block to the DAG, of which $f$ nodes are \seb{possibly faulty or Byzantine}. Every round has at most $n$ blocks (one for each node\footnote{Cordial Miners allows multiple blocks from one faulty node per round. To mitigate equivocating blocks, this protocol suggests the following: once an honest node detects an equivocator, the further equivocator's blocks are not directly referenced by the honest node.}) where every newly created block must refer to $n-f$ blocks from the previous round. In these protocols, each node maintains its local version of the DAG and runs the consensus protocol through logical interpretation of the DAG, i.e., blocks represent transaction proposals and references serve as votes. Therefore, consensus does not require any extra communication complexity. These protocols are leader-based where a leader is elected after a constant number of rounds (wave) using the encoded randomness in each wave in the asynchronous model or the round-robin method in the partially synchronous model. Once the elected leader block is committed and every node performs ordering on the leader block's casual history.



Out of the above protocols, only a few stress fairness while catering to high throughput. Under the assumption of their network model, in these DAG-based protocols, transactions and blocks propagated by the nodes/users in the P2P network might take a long time before a validator node might receive them. Due to that, validator nodes only see a portion of all the users' transactions and blocks ever produced in the network. Therefore, providing fairness to all the issued transactions and blocks is a challenging but important task.

\section{Different notions of Fairness in DLTs} \label{Fairness}

\subsection{Fairness to the Participants}

\begin{enumerate}[leftmargin=*]
    \item \textit{For Validators:} Fairness for the validators refers to having an equal chance for each validator node\footnote{Throughout the paper, we use validator node and node interchangeably unless explicitly specified.} to append its block to a distributed ledger irrespective of communication latency and the validators' computing power. \np{A slow yet honest} validator node cannot be distinguished from an adversarial node, as it becomes challenging to identify the cause of delay in the received block or transaction.

    
Fairness in this prospect has been taken into account in DLT protocols such as Prime~\cite{amir2010prime}, HoneyBadger~\cite{miller2016honey}, and Fairledger~\cite{lev2019fairledger}. These are BFT-based blockchain protocols. 
\seb{To ensure fairness in these protocols, transactions from multiple nodes are consolidated into batches, which are then committed atomically within each epoch. Furthermore, these protocols incorporate mechanisms to identify and penalize nodes that deviate from the prescribed rules, thus bolstering fairness in favor of the honest nodes.}

A recent line of work in DAG-based DLTs starting from Aleph~\cite{gkagol2019aleph}, DAG-Rider~\cite{keidar2021all}, Tusk~\cite{danezis2022narwhal}, and Bullshark~\cite{spiegelman2022bullshark} emphasise on fairness to the validators. During the construction of the DAG in certain DAG-based DLTs, a degree of fairness is attained by mandating the inclusion of a minimum of $n-f$ blocks from different validator nodes in the previous round. To provide fairness for all the validators, some of these protocols~\cite{keidar2021all, spiegelman2022bullshark} use weak links to append slow validators' blocks. The idea of a weak link is to reference a tip  (a block that is not yet referenced) in the distant past together with other ``recent'' tips while creating a new block in the DAG.

Fairness for validators ensures censorship resistance. It also shows a strong notion of participation equity which refers to the Chain Quality property of DLT, meaning that the ledger includes a certain portion of honest validators' blocks. Therefore, to enforce fairness for the validator nodes, the new block should include references to the slow validator's block either by a weak link in structured DAG-based DLTs or by referencing all the current tips.
    
    \item \textit{For Clients:} Transactions in DLTs can be delayed inappropriately by a malicious validator node. A validator node can prevent sending the transactions received from a client to the rest of the DLT network. This particular phenomenon can be termed as a compromise of fair access due to the prevention of a client's transaction from fairly accessing the network relative to other clients' transactions.
    
    Fairness for a client can be defined as fair access to the client's transaction to the rest of the DLT network. It means when a validator node receives a client transaction, the node should send the transaction to the rest of the network without delay. To provide fairness to multiple clients connected to a validator, the validator can either process the clients' transactions in ``First In First Out''(FIFO) order~\cite{duan2014byzid} or by monitoring the latency of the transaction requests~\cite{aublin2013rbft}. 
    
    In protocols like Narwhal~\cite{danezis2022narwhal}, a client transmits its transaction to all the validator nodes to ensure fairness although it increases the bandwidth usage. If the client's transaction is not sequenced in the ledger in time, the client re-submits its transaction. However, re-submitting the transaction to multiple validator nodes not only wastes excessive resources such as bandwidth and CPU but also comes under the radar of request duplication attacks. This attack might harm honest clients if there is a penalty involved in every re-transmission of honest clients' transactions, for example, in Hedera Hashgraph~\cite{baird2019hedera}. 
    To prevent this duplication attack, a technique used in Mir-BFT~\cite{stathakopoulou2019mir} based on hashing can be employed in DAG-based DLTs to provide fairness. On the positive side, the dissemination of transactions is encoded in the very structure of the Hashgraph. Each transaction carries with it the gossip history which is used to achieve consensus on the validity and order of transactions.

A malicious validator node could violate fair access. It can delete or ignore client transactions. It can also delay a client's transaction and favor other clients' transactions in order to punish the client or front-run the client's transaction. With the lack of delay in terms of fair access compromise, the client's transaction is transmitted and received later in the network and hence the transaction is added later in \seb{the ledger} than it should have been.

Clients can achieve fair access to their transactions in different ways as follows:
    \begin{itemize}
        \item Client can run a node that relieves the requirement of middleman proxy for the client in the DLT network, e.g., Permissionless Hedera~\cite{baird2019hedera}.
        \item Client can submit a transaction to multiple or all validators in the DLT network~\cite{danezis2022narwhal}.
        \item Client can submit a transaction to a node it trusts or the node having a better rating if the nodes maintain a reputation system within the DLT network~\cite{huang2020repchain}.
        \item Client can submit a transaction anonymously along with providing proof of correctness to the transaction.
    \end{itemize}

\end{enumerate}

\subsection{Fairness in Consensus Ordering}
Fairness in consensus ordering aka \textit{Order-Fairness} ensures that the order of transactions relative to each other is fair. There are different ways to ensure fairness in transaction ordering. 
Order-Fairness is described in many DLT protocols such as Helix~\cite{asayag2018fair}, Hashgraph~\cite{baird2019hedera}. Kelkar et al.~\cite{kelkar2020order} presents a good overview of order fairness in Byzantine consensus protocols. In a similar direction, Kursawe~\cite{kursawe2020wendy} unveils a group of low overhead protocols, Wendy, which can implement different notions of order-fairness in the blockchain. Another interesting work~\cite{malkhi2022maximal} manifests a notion of content-agnostic order-fairness.

Fairness in ordering can be abused by recording the transaction inappropriately, meaning the transaction that arrived first is recorded as received later. In other words, it is referred to as order manipulation. Particularly in a leader-based consensus of a DLT, the consensus algorithm elects a leader who is responsible for ordering and can delay the transaction of a specific client while ordering. This phenomenon of manipulation of transaction ordering is called the Miner Extractable Value (MEV) where a validator (leader/miner) maximizes its profit by including, excluding, or changing the order of client's transactions within blocks. MEV was first introduced as a measure in Flash Boys 2.0~\cite{daian2020flash}, later it was named as Maximal Extractable Value.

Fair ordering of transactions can solve the problem of front-running attacks 
and censorship resistance in DLTs. The need for the fair ordering of transactions becomes clear when we look at financial systems such as Ethereum~\cite{wood2014ethereum}. A recent report~\cite{barczentewicz2023mev} delineates a comprehensive study of MEV extraction on Ethereum and an overview of policymakers on MEV.  Another latest paper~\cite{yang2022sok} presents a Systematization of Knowledge (SoK) on MEV countermeasures where a brief description of common types of MEV and transaction ordering methods are shown in a precise way.

In the pursuit of achieving order-fairness and mitigating Miner Extractable Value (MEV) in DLT systems, two primary defensive measures are commonly employed: 1) Time-based Order-Fairness and 2) Blind Order-Fairness.

\subsubsection{Time-based Order-Fairness}
This protects from order manipulation by defining certain properties about transaction ordering based on transaction arrival/sent time, and by satisfying these properties. There are multiple definitions for Time-based Order-Fairness in the literature~\cite{kursawe2020wendy,kelkar2020order}. From these literature, following we exhibit a few informal definitions of this order-fairness. Let $N$ represent the total number of validator nodes, and $\alpha$ is a fraction parameter denoting enough number of validator nodes.

\begin{itemize}[leftmargin=*]
    \item \textbf{Receive-Order-Fairness}
        \begin{definition}{(Receive-Order-Fairness)}
            It states that if a sufficiently large number ${\alpha}N$ \seb{(with $\alpha>0.5$)} of validator nodes receive a transaction $tx_1$ before a transaction $tx_2$, then $tx_1$ should be ordered before $tx_2$ in the final ordering of transactions agreed upon by the validator nodes.
        \end{definition}

        For the above definition to work, we need to make a few assumptions: 1) The network must be synchronous; 2) The adversary must be non-colluding and non-corrupting. Otherwise, it is almost impossible to achieve receive-order-fairness. In some sense, \seb{this describes a distributed} FIFO notion.
    \item \textbf{Send-Order-Fairness}
        \begin{definition}{(Send-Order-Fairness)}
            It states that if a transaction $tx_1$ is sent before a transaction $tx_2$, then $tx_1$ should be ordered before $tx_2$ in the agreed-upon transaction log by the validator nodes.
        \end{definition}
        For the above definition to work, we need a trusted way to timestamp the transaction at the client's end. One potential way to generate a trusted timestamp is by using Trusted Execution Environment (TEE) such as Intel SGX~\cite{costan2016intel} or ARM Trustzone~\cite{pinto2019demystifying} at the client side. Nevertheless, even after having a timestamp mechanism, network synchrony still plays a role to ensure that the transaction is not arbitrarily delayed.
    \item \textbf{Approximate-Order-Fairness}
        \begin{definition}{(Approximate-Order-Fairness)} 
            For any two transactions $tx_1$ and $tx_2$, let $n$ validator nodes receive both transactions in a given time-frame and let $\epsilon$ be the number of rounds $tx_1$ and $tx_2$ are apart. Approximate-Order-Fairness states that if ${\alpha}n$ nodes receive $tx_1$ more than $\epsilon$ rounds before $tx_2$, then for every honest node $j$, $j$ does not deliver $tx_2$, unless it has previously delivered $tx_1$.
        \end{definition}
    \item \textbf{Block-Order-Fairness}
        \begin{definition}{(Block-Order-Fairness)}
            For any two transactions $tx_1$ and $tx_2$, let $n$ validator nodes receive both transactions in a given time-frame and let ${\alpha} > \frac{1}{2}$. Block-Order-Fairness states that if at least ${\alpha}n$ nodes receive $tx_1$ before $tx_2$, then for every honest node $j$, $j$ does not deliver $tx_1$ in a later block than it delivers $tx_2$.
        \end{definition}
\end{itemize}

The last two definitions are weaker notions of order-fairness. These two definitions have inherent limitations related to the network model and relative measures. To overcome these limitations, Cachin et al.~\cite{cachin2022quick} presented a new notion of order-fairness, which they call \textit{Differential Order-Fairness}.

Although most of the above definitions are used in the context of Byzantine consensus, these definitions can also be used to provide order-fairness in DAG-based DLT protocols.
 
\subsubsection{Blind Order-Fairness} This provides MEV protection by committing to a transaction ordering without seeing the content of the transactions. That means this type of ordering is determined independently of transaction content. It is also referred to as content-agnostic ordering. The natural way of performing such ordering is by realizing through commit-reveal protocols. 

The idea of a commit-reveal protocol for ordering is to first commit the transactions and reveal the committed transactions once the ordering has already been determined. Nevertheless, there is an inherent leakage of metadata in this ordering which makes the ordering weaker compared to time-based ordering. Helix protocol~\cite{asayag2018fair} employs commit-reveal to provide fair order. Although Helix provides censorship resistance it suffers from order manipulation due to metadata leakage.

The commit-reveal is performed by clients using some cryptographic primitive. Clients send their encrypted transactions as commitments along with metadata to the validator nodes. Then the consensus protocol commits to an ordering of transactions. Later, the commitments are opened and the transactions are executed. The commit-reveal step can be instantiated using different cryptographic primitives. Heimbach and Wattenhofer~\cite{heimbach2022sok} present a Systematization of Knowledge of commit-reveal approaches using different cryptographic primitives in on-chain and off-chain. These approaches can also be helpful in DLT to perform the commit-reveal step. We present a brief overview of some of these primitives and their usage in DLTs to achieve blind order-fairness as follows.

\begin{itemize}[leftmargin=*]
    \item \textbf{Threshold Encryption} Threshold encryption allows a threshold number of honest validator nodes to retrieve committed transactions. In a threshold encryption scheme, during setup, the validator nodes generate a public key, and the corresponding secret key is shared among validator nodes. Using this scheme, a client encrypts a transaction using a public key in the commit phase, later in the reveal phase the message is decrypted using the decryption shares from a threshold number of validator nodes. 

    Malkhi and Szalachowski~\cite{malkhi2022maximal} construct a DAG-based protocol Fino which employs a threshold encryption scheme to achieve blind order-fairness. They also present other approaches, i.e., verifiable secret sharing scheme~\cite{shamir1979share}, AVID-M~\cite{zhang2020byzantine} to achieve blind order-fairness in DAG-based DLTs.

    \item \textbf{Trusted Execution Environment (TEE)} A TEE is a hardware-protected secure area that provides an isolated execution environment for code and data. Hence, it provides confidentiality and integrity of code and data. The state-of-the-art popular implementation of TEE includes Intel SGX~\cite{costan2016intel} and ARM Trustzone~\cite{pinto2019demystifying}. 

    To provide blind order-fairness, for a round, a client first generates a key pair inside TEE and encrypts its transaction using the public key. TEE can be programmed in a way such that the decryption key for the round is released only after the ordering for the round has already been done. TEE-employed order-fairness can be used in the DAG-based DLT such as Teegraph~\cite{fu2022teegraph} which is already designed using TEE.

    \item \textbf{Time-lock Puzzle} Time-lock puzzle~\cite{rivest1996time} can be used to encrypt the transaction as a commitment and the decryption only takes place after enough time has elapsed. For the time-lock puzzle to work in the ordering protocol, the global time-lock parameters must be set so that only after the transaction ordering, the encrypted transactions can be decrypted. Moreover, an adversary should not be able to solve a time-lock puzzle faster than an honest client.
    
Another cryptographic primitive related to the time-lock puzzle is delay encryption~\cite{burdges2021delay} where all clients encrypt their transactions under the same key for a given round. Delay encryption has been used to provide fairness in FairPoS consensus protocol~\cite{chiang2022fairpos}. 
\seb{These primitives have the potential to be used in any DLTs.}
   
\end{itemize}

\subsection{Fairness in terms of DLT Components}
\begin{enumerate}[leftmargin=*]
    \item \textit{Block-level Fairness:} 
Achieving block-level fairness can be approached through various methods. One approach involves establishing weak links that connect the blocks of slow yet honest validators during the creation of new blocks in the DLT.
Attaining block-level fairness is challenging. However, some degree of fairness can be achieved through the tip selection mechanism. If the mechanism selects a tip from a distant past, there is a higher likelihood of including the block from a slow honest validator in the ledger. However, achieving this level of fairness necessitates validator nodes to possess substantial storage capacity to store the complete DLT history, which can be practically challenging to implement without any relaxation. Additionally, discarding old blocks poses difficulties, as it may inadvertently remove blocks from slow validators that were never added to the ledger. DAG-Rider and Bullshark are protocols that employ weak links to achieve block-level fairness.
    
    \item \textit{Transaction-level Fairness:} 
This principle asserts that transactions received by an honest validator node will eventually be included in the ledger. Transaction-level fairness is particularly relevant for ensuring censorship resistance and is considered a more realistic measure than block-level fairness. 
To achieve transaction-level fairness, it is crucial for honest validator nodes to promptly relay client transactions to the DLT network upon receiving them. However, due to network connectivity issues, such as those encountered by slow validator nodes, transactions or blocks may not reach all nodes in the network.
In such cases, transaction-level fairness can be upheld by re-injecting the transactions from a non-committed block belonging to a slow honest validator into a newly created block. This process ensures that the transactions ultimately find their way into the ledger.
\end{enumerate}

\section{Discussion} \label{Discussion}
\seb{In the ensuing discussion, we delve into various aspects and implications of the findings presented in the preceding sections, shedding light on the broader implications of the research and exploring potential avenues for further investigation.} In a DLT, fairness notions are directly or indirectly affected by different components or functionalities plugged into it. In this section, we elaborate on the relation of some of these components with fairness notions.
\subsection{Writing Access}

Writing access permits nodes to propose and write blocks, hence progressing the state of the ledger. A DLT protocol can pertain to a high degree of parallelism by having concurrent writing access by all the participants. 

The fairness experienced by participants in a system is influenced by the writing access policies in place. In transaction-based DAG ledgers, where nodes add transactions to the ledger, providing writing access to all nodes ensures fairness among them. On the other hand, block-based DAG ledgers involve different node roles (validator, miner, ordinary, follower), and only a subset of nodes (such as miners) possess the permission to append blocks. This restricted access can introduce fairness limitations to the system. Thus, a more open writing access approach leads to a fairer system overall.
Currently, the prevailing writing access solutions in the field encompass PoW and PoS-based lotteries, alongside the permissioned setting. 

Let us mention a
few works that allow more general writing access. Zhao et al.~\cite{zhao2021secure} compose an improved access control mechanism for DAG-based DLTs to improve the security and robustness of the network. The work focuses on fairness in the aspect of network parameters such as fairness in dissemination rate and latency. Cullen et al.~\cite{cullen2021access} achieve fairness in an adversarial environment by proposing an access control mechanism for DAG-based DLTs. They ensure fairness during congestion using a notion of weighted max-min fairness that shows each node gets a fair share of throughput weighted by the reputation score of each node.  

\subsection{Garbage Collection}
In DAG-based DLT protocols, nodes require unbounded memory to guarantee validity and fairness. This arises the practical challenges for these protocols to be deployed. Due to the requirement of unbounded memory, nodes are not allowed to garbage collect the old rounds. Otherwise, an honest slow node block might be garbage collected before being added to the DAG. As per this scenario, garbage collection directly conflicts with fairness, especially in an asynchronous network model where a block or message can be delayed for an indefinite time. Garbage collection is a trade-off between fairness and performance.

Narwhal~\cite{danezis2022narwhal} employs a garbage collection mechanism to clean the DAG up to a certain round in the past from the genesis. Nevertheless, fairness at the block-level is sacrificed because slow nodes might be garbage collected before they get a chance to be ordered. Bullshark~\cite{spiegelman2022bullshark} on the other hand, achieves fairness for all the nodes in an asynchronous network while having a garbage collection mechanism in place. Bullshark gives a practical implementation where nodes have bounded memory and fairness is achieved after Global Stabilization Time (GST) during the period of synchrony. 

The garbage collection mechanism is rather new in the DAG-based DLT space. Mechanisms similar to the one in Bullshark can be adopted in other DAG-based DLTs. \seb{This kind of mechanism is also required for any high-throughput DLT since eventually, the underlying data structure has to be pruned.}

\subsection{Reward Mechanism} 
In distributed ledgers, the reward (incentive) mechanism plays a crucial role to ensure fairness among the participants. The reward mechanism should reward the nodes according to their effort put forward to make the protocol progress. 

In the majority of DLTs, expected rewards are typically proportional to the delivered Proof-of-Work (PoW) or Proof-of-Stake (PoS), as well as the performance of the nodes, resulting in a fair distribution. However, as a second-order effect, nodes with higher PoW or PoS tend to exhibit less variance in their profits and incur lower costs in running their nodes. This phenomenon can lead to centralization, where a small number of nodes with significant resources dominate the network.

To address this potential centralization, many PoS systems impose a cap on the maximum amount of staked funds. This limitation aims to create a more decentralized and fairer distribution of rewards. 

Several DAG-based DLTs do not have a reward mechanism as being ``feeless'' is an advantage for those DAG-based DLTs, e.g., \cite{popov2018tangle}. Moreover, a few DAG-based DLTs e.g., Inclusive~\cite{lewenberg2015inclusive}, Sphinx~\cite{wang2021weak} employ reward mechanisms to increase network participation, and Graphchain~\cite{boyen2018graphchain} introduces its reward mechanism to maintain the DAG. 

Some DLT protocols do not uniformly distribute the rewards and hence lack fairness. For example, 
scalable BFT protocols~\cite{gueta1804sbft} employing threshold signature distribute the rewards to every node no matter whether the node participated or not. Nonetheless, Kokoris-Kogias~\cite{kokoris2019robust} builds a fair reward mechanism in the novel design of robust and scalable BFT protocol which can be deployed in BFT-based DLT.

In SUI protocol~\cite{sui_2022_tokenomics}, which is based on Bullshark~\cite{spiegelman2022bullshark},  fairness is achieved by rewarding validators based on their behavior and performance, as perceived by other validators. This is done through subjective reporting which is then used as input into the stake reward distribution rule. 


\subsection{Tip Selection}
In the Nakamoto consensus mechanism~\cite{nakamoto2008bitcoin}, each block refers to a single previous block, resulting in an ever-extending chain. Conversely, DAG-based systems allow blocks to reference multiple preceding blocks, leading to a more intricate structure as opposed to a straightforward, linear chain. As a result of this partial order of vertices in a constructed DAG, the tip selection algorithm becomes an integral part of the associated consensus mechanism. Fairness in the context of tip selection algorithms generally refers to an equal opportunity for all transactions to get selected and validated regardless of their origin, ensuring that no transaction gets left out. We list a few tip selection algorithms:
\begin{itemize}[leftmargin=*]
    \item \textbf{Uniform Random Tip Selection}  To issue a new block, a node chooses tips to approve uniformly at random from all \textit{good} tips in the DAG until
 a fixed number $k$ of references is created, e.g., IOTA~\cite{popov2020coordicide,muller2022tangle}. This algorithm is specifically designed for synchronous network models and permissionless DAG-based DLTs since the size of the references should be negligible compared to the block size. A good tip refers to a tip that is valid and consistent with the finalized ledger and issued rather recently. Such tip selection gives a fair chance to all good tips in the tip pool to be referenced and eventually approved.  Moreover, it leads to a Nash Equilibrium, i.e., knowing that other
nodes select tips with equal probabilities, a given node has nothing to gain from deviating from this behavior, \seb{see \cite{EquilibriumTangle}.}
 \item \textbf{Round-based Tip Selection} To issue a new block at round $r$, a node chooses at least $n-f$ tips in the DAG from round $r-1$, e.g., Aleph~\cite{gkagol2019aleph}, Bullshark~\cite{spiegelman2022bullshark}, Cordial Miners~\cite{keidar2022cordial}, DAG-Rider~\cite{keidar2021all}, Tusk~\cite{danezis2022narwhal}. Such round-based tip selection algorithms are used in permissioned DAG-based DLT protocols which work under partially synchronous or asynchronous network models. The number $n-f$ is the maximum guaranteed number of tips one can expect if $f$ nodes are faulty or Byzantine. Moreover, certain protocols~\cite{spiegelman2022bullshark,keidar2021all}, offer some sort of fairness to slower nodes. This is achieved by allowing nodes to reference blocks from previous rounds using weak links. These weak links facilitate the inclusion of transactions from slower nodes into the ledger, while not being utilized for voting purposes.
 \item \textbf{Heaviest Cluster Tip Selection} To issue a new block, a node selects all tips from the heaviest cluster, the largest subset of blocks in the DAG, in which each block is not comparable by the partial order with at most a constant number of other blocks, e.g.,~Phantom~\cite{sompolinsky2018phantom}, DAG-KNIGHT~\cite{sompolinsky2022dag}. This approach achieves certain fairness as a new block references all existing tips in the heaviest subDAG.
 \item \textbf{Sub-sampled Voting Tip Selection} To issue a new block, a node first queries small random samples of other nodes and based on their responses finds the tips that are likely to be approved by the network, e.g., \cite{rocket2019scalable} and \cite{popov2020coordicide}. Due to its nature, the algorithm works with permissionless DAG-based DLT protocols. This tip selection algorithm is considered fair as with a high probability it reflects the preference of the majority of the network. 
 \end{itemize}

\subsection{Smart Contract Architecture}
The increased level of parallelism in DAG-based DLTs shows promise in addressing contention issues within UTXO, Cardano, and object-based smart contracts like SUI~\cite{sui_2022_wp}. These smart contracts offer distinct advantages compared to account-based smart contracts, particularly in terms of determinism. The deterministic nature of these smart contracts significantly reduces the occurrence of Miner Extractable Value (MEV) possibilities.

The inherent deterministic execution model and parallelism in object-based smart contract systems provide notable advantages in terms of fairness, scalability, and security. 
The deterministic nature of object-based smart contracts ensures predictable outcomes, facilitating auditing and verification processes. Overall, these advantages position object-based smart contracts as an attractive option for achieving fairness and performance in decentralized applications within the context of DAG-based DLTs.
\section{Conclusion} \label{Conclusion}

In this study, we have conducted a thorough analysis of fairness in DAG-based DLTs, defining different notions and examining their entanglement with DLT components. Promising research directions include exploring alternative methods for fairness without weak references and investigating cryptographic schemes for blind order-fairness. By addressing these areas, we can enhance the understanding and implementation of fairness in DAG-based DLTs, fostering more equitable and robust systems.

\section{Acknowledgement}
Mayank Raikwar has been supported by IOTA Ecosystem Development grant.

\bibliographystyle{IEEEtran}
\bibliography{report}

\end{document}